\documentclass[]{article}
\usepackage{amsmath}
\usepackage{amssymb}
\usepackage{graphicx}

\newcommand{\verteq}[0]{\scriptscriptstyle\|}

\title{Second-Order, Dissipative T\^{a}tonnement: Economic Interpretation and 2-Point Limit Cycles}
\author{Eric Kemp-Benedict\\
Stockholm Environment Institute\\
eric.kemp-benedict@sei-international.org}

\begin{document}

\maketitle
\bibliographystyle{plain}

\begin{abstract}
This paper proposes an alternative to the classical price-adjustment mechanism (called ``t\^{a}tonnement'' after Walras) that is second-order in time. The proposed mechanism, an analogue to the damped harmonic oscillator, provides a dynamic equilibration process that depends only on local information. We show how such a process can result from simple behavioural rules. The discrete-time form of the model can result in two-step limit cycles, but as the distance covered by the cycle depends on the size of the damping, the proposed mechanism can lead to both highly unstable and relatively stable behaviour, as observed in real economies.

\vspace{2em}

\end{abstract}

\section{Introduction}
Economic life is plagued by uncertainty; speculation, crop failures, and new entries into a market can shift prices, sometimes dramatically. Economic theory reflects this uncertainty, and in the search for market equilibria economists found that instability can arise from small changes in behaviour in an already stable economy. However, these results appear to have gone too far \cite{ackerman_still_2001}, because they imply both that instability should be a common occurrence \cite{scarf_examples_1960} and that an already stable economy can become unstable after modest changes in behaviour \cite{kirman_market_1986}. Moreover, the discrete-time version of dynamic price models can be unstable, even if the continuous-time version is stable \cite{saari_iterative_1985}; and discrete-time dynamics are more relevant to economics \cite{kitti_iterative_2004}. This degree of model instability is inconsistent with much economic activity because, outside of specialized markets like the stock market or real estate, most people do not experience extreme changes in prices in their day-to-day lives. They go to the local market or store with the expectation that prices may have shifted a bit, but not wildly, giving them enough certainty to plan and pace their spending.\footnote{From personal experience, this is true even in bazaars, where bargaining is the norm. Sellers know the ``correct'' price, and knowledgeable buyers know the standard markup.}

Instabilities in economic theory arise from two sources: lack of constraints on the aggregate response to a change in prices, and the price adjustment dynamic. The classical price adjustment mechanism, called ``t\^{a}tonnement'', or ``groping'', by L\'eon Walras \cite{walras_elements_1954, walker_walrass_1987} is a first-order dynamic equation, as formulated by Samuelson \cite{samuelson_foundations_1947}. In this paper we focus on a particular type of instability, non-convergence in economies with stable equilibria---that is, economies where the Jacobian of the excess demand function has negative eigenvalues at one or more equilibrium. We argue, by analogy with the damped harmonic oscillator, that a second-order discrete-time equation with damping can capture both the instability (e.g., the stock market) and relative stability (e.g., at the local market) seen in real economies. The difference is caused by the coefficient of the damping term. We show how the proposed mechanism can arise from simple behavioural rules.

\subsection{The SMD theorem}
Economic statics and dynamics derive from behavioural responses to relative supply and demand of different commodities in a market. The behaviour of both sellers and buyers is stimulated by the market price of the commodity. Economic statics finds the price that equates supply and demand, while economic dynamics specifies how prices change over time when supplies and demands differ. When supplies and demands are equal for all commodities---that is, excess demand, the difference between demand and supply, is zero for every commodity---the system is at equilibrium, and the time rate of change of every price is zero.

Excess demand is captured by the excess demand function, $\mathbf{\xi}(\mathbf{p})$. It is a vector function, with a value for each commodity, that depends on a vector of prices, one for each commodity. In the theory of General Equilibrium, the excess demand function has the following properties \cite{ginsburgh_structure_2002}: i) it is continuous; ii) it is homogeneous of degree zero; and, iii) it satisfies Walras' law, $\mathbf{p}\cdot\mathbf{\xi}(\mathbf{p}) = 0$. Walras' law says that, for the economy as a whole, the money value of total demand must equal the money value of total supply, although individual markets can be out of equilibrium.

In a series of paper, Sonnenschein \cite{sonnenschein_market_1972}, Mantel \cite{mantel_characterization_1974}, and Debreu \cite{debreu_excess_1974} showed that any function satisfying the three conditions listed above can be expressed as a sum of well-behaved individual demand functions. This theorem, called the ``SMD theorem'' after the three authors, was a startling result, because it implies that microeconomic foundations place almost no constraints on the possible forms of the macroeconomy. Later, Kirman and Koch showed that the SMD result holds even if consumers have almost identical preferences \cite{kirman_market_1986}. The general conclusions of the theorem have been shown to be quite robust \cite{ackerman_still_2001}.

\subsection{Classical t\^{a}tonnement and the ubiquity of instability}
The SMD theorem by itself says nothing about stability of the general equilibrium. Any statement about stability requires dynamics---a mathematical statement of how prices change. Following Walras and Samuelson, classical t\^atonnement is a first-order dynamic process,
\begin{equation}\label{class_taton_cont}
\frac{d\mathbf{p}}{d t} = k|\mathbf{p}|\mathbf{\xi}(\mathbf{p}).
\end{equation}
In this paper we assume that the prices of all commodities respond in the same way to excess demand, and that demand for all goods is expressed in a comparable unit, $[\mathrm{demand}]$; we do not have to make these assumptions, but it simplifies the presentation. In the equation above, the factor of $|\mathbf{p}|$ ensures that when the price vector is scaled by an arbitrary factor, both sides of the equation remain in balance. The coefficient $k$, which is assumed to be the same for all commodities, has units of $[\mathrm{demand}]^{-1}[T]^{-1}$.

In a discrete-time representation, which is more realistic because reaction to a change in prices is delayed from one time interval to the next \cite{kitti_iterative_2004}, this dynamic can be expressed as
\begin{equation}\label{class_taton_discr}
\mathbf{p}_{N+1} = \mathbf{p}_N + k|\mathbf{p}|\mathbf{\xi}(\mathbf{p}_N) \Delta t.
\end{equation}
When excess demand is high, the price rises proportional to the gap. When it is low, the price falls, again proportional to the gap. Scarf \cite{scarf_examples_1960} showed examples of excess demand functions, expressed as a sum of reasonable individual demand functions, that have no stable equilibria for any set of prices under traditional t\^atonnement. More importantly for this paper, he showed examples of repetitive limit cycles. All of his limit cycles arose in theories with unstable equilibria. Later, Saari \cite{saari_effective_1978,saari_iterative_1985} showed that any price adjustment mechanism that involves the first difference of the price (but, possibly, historical values of the excess demand function) will, for some excess demand functions, fail to converge from an open set of initial points to a stable equilibrium. Saari considered discrete-time models and showed that even with stable equilibria the dynamic processes that he considered could give rise to limit cycles and other non-convergent trajectories.

Instability is a feature of real economies, so from one view the Saari results are not surprising. However, because of the SMD theorem, a well-functioning economy can, if consumers or producers change their behaviour even slightly, shift to an unstable state in which prices vary wildly; this is a problem. For most people, especially in high-income countries, daily economic life is relatively stable. Also, limit cycles are rare; except for seasonal price changes, which are driven by external, not internal, forces, sellers do not repeatedly cycle through a set of prices.

\subsection{Alternatives to classical t\^{a}tonnement}
Failure to converge is often identified with the SMD theorem. It is indeed a consequence of the SMD result, but also depends on economic dynamics. Classical t\^{a}tonnement, with its first-order dynamics, seems like a parsimonious model, but if it means widespread instability, then it does not agree with much of day-to-day economic life. Saari \cite{saari_iterative_1985} studied a general class of iterative price mechanisms, but each of them was first-order (or first-difference) in price. Subsequent authors have identified a variety of unstable and chaotic behaviour for different iterative price mechanisms \cite{bala_chaotic_1992, tuinstra_price_1997,weddepohl_simulating_1997}, while others have identified mechanisms that reduce instability \cite{bala_note_1998,goeree_stability_1998,kitti_iterative_2004,cole_fast-converging_2008,kitti_convergence_2010} by either restricting the form of the excess demand function, bounding changes in price, or both. However, each of these proposals uses a first-difference dynamic, determining prices at one time step from prices at the immediately previous time step.

In this paper we propose an alternative dynamic, suggested by the damped harmonic oscillator. It allows for quasi-periodic behaviour but also dissipation, so that disturbances away from equilibrium gradually decrease.

\section{A second-order t\^{a}tonnement process}
The classical t\^{a}tonnement process in Equations (\ref{class_taton_cont}, \ref{class_taton_discr}) has a clear economic interpretation: if demand is higher than supply, sellers adjust prices upward by an amount proportional to the gap; if demand is low, they adjust prices downward to try to stimulate demand. In the second-order process proposed in this paper, the \textit{rate of change of the rate of change of price} responds to non-zero excess demand. When a gap opens between demand and supply, sellers do not respond right away. Instead, they start to increase prices, gradually at first, and then, if the excess demand gap persists, more rapidly.

To motivate the mechanism, we show how the basic behaviour emerges from a simple model in which there are two types of sellers: group $a$ watches sales and inventory to gauge the level of excess demand and set its price, while group $b$ looks at what other people are doing, setting their own price increase at time $t$ based on the average price rise at time $t-1$. As we now show, these rules give rise to a second-order dynamic that can lead to cyclic, or ``pulsing'' behaviour, as observed in experimental situations {\cite{ostrom_rules_1994}}. These rules can be written\footnote{Properly, we should use the mean of excess demand evaluated at different prices, rather than excess demand at the mean price. However, since we are motivating the formula rather than deriving it, we assume for simplicity that all price movements are such that $\overline{{\mathbf\xi}(\mathbf{p}_{t-1})}\approxeq \mathbf{\xi}(\bar{\mathbf{p}}_{t-1})$.} 
\begin{subequations}\label{simple_model}
\begin{align}
\Delta {\mathbf p}_{a,t} &= \mu {\mathbf\xi}(\bar{\mathbf p}_{t-1}),\label{simple_model_a}\\
\Delta {\mathbf p}_{b,t} &= \nu\Delta{\bar {\mathbf p}}_{t-1}.\label{simple_model_b}
\end{align}
\end{subequations}
If a fraction $f_a$ of sellers are of type $a$, and a fraction $f_b = 1-f_a$ are of type $b$, then we can compute $\Delta\bar{\mathbf p}_t = f_a \Delta {\mathbf p}_{a,t} + f_b \Delta {\mathbf p}_{b,t}$ to give
\begin{equation}\label{simple_model_ave}
\Delta\bar{\mathbf p}_t = f_a\mu {\mathbf\xi}(\bar{\mathbf p}_{t-1}) + f_b\nu\Delta{\bar {\mathbf p}}_{t-1}.
\end{equation}
Subtracting $\Delta{\bar {\mathbf p}}_{t-1}$ from both sides gives a second-order equation,
\begin{equation}\label{simple_model_secorder}
\Delta\bar{\mathbf p}_t - \Delta{\bar {\mathbf p}}_{t-1} = f_a\mu {\mathbf\xi}(\bar{\mathbf p}_{t-1}) - (1 - f_b\nu)\Delta{\bar {\mathbf p}}_{t-1}.
\end{equation}
if $f_b\nu < 1$ then the second term on the right-hand side is a damping term, as we show below.

We now present a second-order t\^{a}tonnement process and show some of its properties. We begin by commenting on the mathematics of classical, first-order t\^{a}tonnement.

\subsection{Continuous-time formulation}
The continuous-time version of classical t\^{a}tonnement is shown in Equation (\ref{class_taton_cont}). Suppose that, with this dynamic, an economy starts close to an equilibrium $\mathbf{p}^{*}$, so that $\mathbf{\xi}(\mathbf{p}^{*}) = \mathbf{0}$. Then, because the excess demand function is continuous, we can expand it to first order in the vicinity of the equilibrium, to find
\begin{equation}\label{class_taton_perturb}
\frac{d\Delta \mathbf{p}}{d t} \approx k|\mathbf{p}|D\mathbf{\xi}(\mathbf{p}^{*})\cdot\Delta \mathbf{p},
\end{equation}
where $D\mathbf{\xi}$ is the Jacobian of the excess demand function. Whether an equilibrium is stable depends on the signs of the eigenvalues of the Jacobian, but the homogeneity property of the excess demand function adds a complication. Because the excess demand function is homogeneous of degree zero it always has a zero eigenvalue, proportional to $\mathbf{p}^{*}$. To see this, compute the excess demand function at a point $(1+\epsilon)\mathbf{p}^{*}$ that is close to the equilibrium point. This is simply a scaling of the price vector by $(1+\epsilon)$, so it does not change the value of the excess demand function. Therefore,
\begin{equation}\label{pstar_is_zero_eigenvector}
D\mathbf{\xi}(\mathbf{p}^{*})\cdot\mathbf{p}^{*} = 0.
\end{equation}
A zero mode signals a potential instability, but it is not a problem in this case because, from Walras' law, price dynamics in classical t\^{a}tonnement are restricted to the hyphersphere defined by $|\mathbf{p}|^2=\mathrm{const.}$:
\begin{subequations}\label{class_taton_sphere_restrict}
\begin{align}
\mathbf{p}\cdot\frac{d\mathbf{p}}{d t} &= k|\mathbf{p}|\mathbf{p}\cdot\mathbf{\xi}(\mathbf{p})\label{class_taton_sphere_restrict_a}\\
\frac{1}{2}\frac{d|\mathbf{p}|^2}{d t} &= 0.\label{class_taton_sphere_restrict_b}
\end{align}
\end{subequations}
We note as an aside that it is common practice to restrict the price arbitrarily to a simplex defined by $\sum_{i=1}^{N}p_i = 1$, where $N$ is the number of commodities \cite{ginsburgh_structure_2002}. Equation (\ref{class_taton_sphere_restrict}) shows that re-scaling to the simplex is not necessary, because the dynamics already restrict prices to the hypersphere defined by $\sum_{i=1}^{N}p_i^2 = \mathrm{const.}$ Admittedly, this result only follows if all prices respond in the same way to excess demand, as we assume in Equation (\ref{class_taton_cont}), but a similar result holds if the coefficient $k$ varies by commodity.

Because of the zero mode in the Jacobian, we are interested in the eigenvalues of the eigenvectors that span the space perpendicular to the equilibrium price vector $\mathbf{p}^{*}$. In this paper we are interested in cyclic behaviour around stable equilibria. So, suppose that all of the eigenvalues are negative, and denote the eigenvalue with the smallest absolute value by $-\lambda_m/|\mathbf{p}|$. (The factor of $|\mathbf{p}|$ in the denominator normalizes the eigenvalues so they have units of demand.) Then, from Equation (\ref{class_taton_perturb}), the price disturbance $\Delta\mathbf{p}$ will decay to zero exponentially at least as fast as $\exp(-k\lambda_m t)$.

We now propose the following second-order t\^{a}tonnement process,
\begin{equation}\label{secorder_taton_cont}
\frac{d^2\mathbf{p}}{d t^2} + \mathbf{p}\frac{|\dot{\mathbf{p}}|^2}{|\mathbf{p}|^2} = k|\mathbf{p}|\mathbf{\xi}(\mathbf{p}) - \gamma\frac{d\mathbf{p}}{d t},
\end{equation}
where $\gamma$ is the damping coefficient and $\dot{\mathbf{p}}$ is the time rate of change of the price vector. The somewhat odd second term on the left-hand side of this equation ensures that the price vector stays on the hypersphere $|\mathbf{p}|^2=\mathrm{const.}$ As above, this is necessary to avoid problems with the zero mode parallel to the price vector, shown in Equation (\ref{pstar_is_zero_eigenvector}), that arises from the homogeneity of the excess demand function. To see that the price vector stays on the hypersphere, dot-multiply both sides of the equation by $\mathbf{p}$,
\begin{subequations}\label{secorder_taton_walras_law}
\begin{align}
\mathbf{p}\cdot\left(\frac{d^2\mathbf{p}}{d t^2} + \mathbf{p}\frac{|\dot{\mathbf{p}}|^2}{|\mathbf{p}|^2}\right) &= \mathbf{p}\cdot\left(k|\mathbf{p}|\mathbf{\xi}(\mathbf{p}) - \gamma\frac{d\mathbf{p}}{d t}\right)\label{secorder_taton_walras_law_a}\\
\frac{1}{2}\frac{d^2|\mathbf{p}|^2}{d t^2} - |\dot{\mathbf{p}}|^2 + |\mathbf{p}|^2\frac{|\dot{\mathbf{p}}|^2}{|\mathbf{p}|^2} &= -\frac{1}{2}\gamma\frac{d|\mathbf{p}|^2}{d t}\label{secorder_taton_walras_law_b}\\
\frac{1}{2}\frac{d^2|\mathbf{p}|^2}{d t^2} &= -\frac{1}{2}\gamma\frac{d|\mathbf{p}|^2}{d t}.\label{secorder_taton_walras_law_c}
\end{align}
\end{subequations}
If the magnitude of the price vector is not changing initially, then integrating the final equation shows that it remains unchanging.

The perturbative version of Equation (6), analogous to Equation (\ref{class_taton_perturb}), is
\begin{equation}\label{secorder_taton_perturb}
\frac{d^2\Delta\mathbf{p}}{d t^2} \approx k|\mathbf{p}|D\mathbf{\xi}(\mathbf{p}^{*})\cdot\Delta\mathbf{p} - \gamma\frac{d\Delta\mathbf{p}}{d t}.
\end{equation}
The second term in Equation (\ref{secorder_taton_cont}) is quadratic in $\Delta\mathbf{p}$, so it does not appear in the perturbative equation. This is the damped harmonic oscillator, and the solutions are well known. If all modes of the Jacobian are negative, with the slowest mode equal to $-\lambda_m/|\mathbf{p}|$, then perturbations decay at least as fast as
\begin{equation}\label{secorder_taton_perturb_solns}
r = -\frac{\gamma}{2} + \sqrt{\left(\frac{\gamma}{2}\right)^2-k\lambda_m}.
\end{equation}
The damping term guarantees that a perturbation away from a stable equilibrium will return to the equilibrium point.

\subsection{Discrete-time formulation}
\begin{figure}[h]
\centering
\includegraphics[scale=0.5]{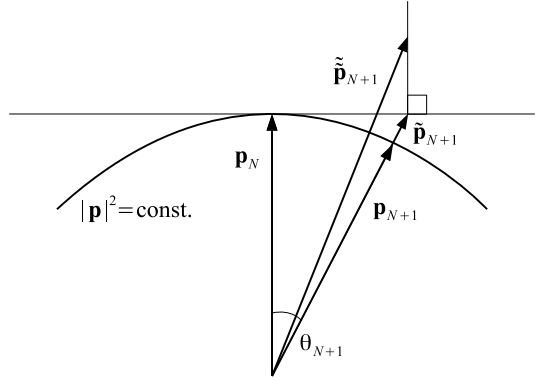}
\caption{Construction of discrete-time price vector}
\label{disc_time_const}
\end{figure}

The discrete-time version of a well-behaved, convergent, continuous process may not converge \cite{saari_iterative_1985}. Discrete-time dynamics are more relevant than continuous-time dynamics to economics, because prices are updated in discrete, sequential events \cite{kitti_iterative_2004}. We therefore propose a discrete-time formulation of the second-order process. In the discrete-time case, avoiding the zero-mode of the excess demand function's Jacobian and staying on the $|\mathbf{p}|^2={\mathrm{const.}}$ hypersphere is more complicated than in the continuous-time case. We therefore develop the discrete-time formulation in stages, starting with a naive version,
\begin{equation}\label{secorder_taton_discr_naive}
\frac{\tilde{\tilde{\mathbf{p}}}_{N+1} - 2\mathbf{p}_N + \mathbf{p}_{N-1}}{\Delta t^2} = k|\mathbf{p}_N|\mathbf{\xi}(\mathbf{\mathbf{p}_N}) - \gamma\frac{\mathbf{p}_N - \mathbf{p}_{N-1}}{\Delta t}.
\end{equation}
The two tildes over $\tilde{\tilde{\mathbf{p}}}$ will be removed as we move from the naive formulation to the final version. The geometric interpretation of each step is shown in Figure \ref{disc_time_const}.

Reorganizing Equation (\ref{secorder_taton_discr_naive}) and collecting the factors of $\Delta t$ into the excess demand function and the damping factor, by defining $\hat{\mathbf{\xi}}(\mathbf{p}) \equiv \Delta t^2\mathbf{\xi}(\mathbf{p})$ for excess demand and $\hat{\gamma} \equiv \gamma\Delta t$ for damping, we find
\begin{equation}\label{secorder_taton_discr_naive_reorg}
\tilde{\tilde{\mathbf{p}}}_{N+1} - \mathbf{p}_N = (1-\hat{\gamma})((\mathbf{p}_N - \mathbf{p}_{N-1}) + k|\mathbf{p}_N|\hat{\mathbf{\xi}}(\mathbf{\mathbf{p}_N}).
\end{equation}
As seen in Figure \ref{disc_time_const}, this equation can give deviations $\tilde{\tilde{\mathbf{p}}}_{N+1} - \mathbf{p}_N$ that are parallel to $\mathbf{p}_N$, in the direction of the zero mode of the Jacobian of the excess demand function. We therefore project the right-hand side of this equation onto the hyperplane perpendicular to $\mathbf{p}_N$ and define a revised difference equation,
\begin{equation}\label{secorder_taton_discr_project}
\begin{split}
\tilde{\mathbf{p}}_{N+1} - \mathbf{p}_N &= \left ( \mathbf{1} - \frac{\mathbf{p}_N \mathbf{p}_N}{|\mathbf{p}_N|^2} \right)\cdot\left( \tilde{\tilde{\mathbf{p}}}_{N+1} - \mathbf{p}_N\right )\\
&= \left ( \mathbf{1} - \frac{\mathbf{p}_N \mathbf{p}_N}{|\mathbf{p}_N|^2} \right)\cdot \left[(1-\hat{\gamma})(\mathbf{p}_N - \mathbf{p}_{N-1}) + k|\mathbf{p}_N|\hat{\mathbf{\xi}}(\mathbf{\mathbf{p}_N}) \right]\\
&= (1-\hat{\gamma})\left(\mathbf{p}_N\frac{\mathbf{p}_N\cdot\mathbf{p}_{N-1}}{|\mathbf{p}_N|^2} - \mathbf{p}_{N-1} \right) + k|\mathbf{p}_N|\hat{\mathbf{\xi}}(\mathbf{p}_N).
\end{split}
\end{equation}
In this calculation we used Walras' law in passing from the second to the third line. This is the second step in our construction, and so we have removed one of the tildes from the price at time step $N+1$.

Our formulation, when we are finished, will preserve the magnitude of the price vector. Accordingly, we can assume that $|\mathbf{p}_{N}|=|\mathbf{p}_{N-1}|$ and write
\begin{equation}\label{secorder_taton_discr_project_cos}
\tilde{\mathbf{p}}_{N+1} - \mathbf{p}_N = (1-\hat{\gamma})\left(\mathbf{p}_N\cos\theta_N - \mathbf{p}_{N-1} \right) + k|\mathbf{p}_N|\hat{\mathbf{\xi}}(\mathbf{p}_N),
\end{equation}
where $\theta_N$ is the angle between $\mathbf{p}_N$ and $\mathbf{p}_{N-1}$. By construction, $|\tilde{\mathbf{p}}_{N+1}|>|\mathbf{p}_N|$. This follows because we have restricted the difference between the two price vectors to the plane tangent to the hypersphere; therefore $\tilde{\mathbf{p}}_{N+1}$ lies outside the hypersphere (see Figure \ref{disc_time_const}). We find that
\begin{equation}\label{pNplus1_over_pN}
|\tilde{\mathbf{p}}_{N+1}|=|\mathbf{p}_N|\sqrt{1+(1-\hat{\gamma})^2\sin^2\theta_N + k^2|\hat{\mathbf{\xi}}(\mathbf{p}_N)|^2 - 2(1-\hat{\gamma})k\frac{\mathbf{p}_{N-1}}{|\mathbf{p}_N|}\cdot\hat{\mathbf{\xi}}(\mathbf{p}_N)}.
\end{equation}
The final step in the construction of the discrete-time version of the t\^{a}tonnement process is to scale $\tilde{\mathbf{p}}_{N+1}$ to have the same length as $\mathbf{p}_N$. This gives the final version for the price vector $\mathbf{p}_{N+1}$ (with no tilde),
\begin{equation}\label{secorder_taton_discr}
\mathbf{p}_{N+1} = A(\mathbf{p}_N,\mathbf{p}_{N-1};\hat{\gamma}) \left[\mathbf{p}_N + (1-\hat{\gamma})\left(\mathbf{p}_N\cos\theta_N - \mathbf{p}_{N-1} \right) + k|\mathbf{p}_N|\hat{\mathbf{\xi}}(\mathbf{p}_N)\right],
\end{equation}
where $A(\mathbf{p}_N,\mathbf{p}_{N-1};\hat{\gamma})$ is equal to $|\mathbf{p}_N|/|\tilde{\mathbf{p}}_{N+1}|$, the inverse of the square root in Equation (\ref{pNplus1_over_pN}). Note that by construction $A(\mathbf{p}_N,\mathbf{p}_{N-1};\hat{\gamma}) < 1$. In fact, as can be seen in Figure \ref{disc_time_const}, it is the cosine of the angle between $\mathbf{p_N}$ and $\mathbf{p_{N+1}}$, or $\cos\theta_{N+1}$.

\subsection{Two-point limit cycles}
Saari demonstrated that the discrete-time formulation of classical t\^{a}tonnement is not guaranteed to converge by constructing two-point limit cycles \cite{saari_iterative_1985}. We therefore discuss two-point limit cycles under our proposed mechanism; while falling short of a full convergence analysis, it is revealing of how the damping mechanism works. Suppose that there is a two-point limit cycle, in which the price vector alternates between two values, $\mathbf{a}$ and $\mathbf{b}$. That is,
\begin{equation}\label{lim_cycle_prices}
\begin{matrix}
\mathbf{b} & \rightarrow & \mathbf{a} & \rightarrow & \mathbf{b} \\
\verteq & & \verteq & & \verteq \\
\mathbf{p}_{N-1} & & \mathbf{p}_N & & \mathbf{p}_{N+1}.
\end{matrix}
\end{equation}
Also suppose that we have chosen the normalization so that $|\mathbf{a}|=|\mathbf{b}|=1$.
Then,
\begin{equation}\label{lim_cycle_orig}
\mathbf{b} = A(\mathbf{a},\mathbf{b};\hat{\gamma}) \left[\mathbf{a} + (1-\hat{\gamma})\left(\mathbf{a}\cos\theta_N - \mathbf{b} \right) + k\hat{\mathbf{\xi}}(\mathbf{a})\right].
\end{equation}
Next, recall that $A(\mathbf{a},\mathbf{b};\hat{\gamma}) = \cos\theta_{N+1}$. For the two-point limit cycle this is the same as $\cos\theta_N$, because $\theta_{N+1} = -\theta_N$. We write the common value as $\cos\alpha$, so that
\begin{equation}\label{lim_cycle_defalpha}
\mathbf{b} = \cos\alpha \left[\mathbf{a} + (1-\hat{\gamma})\left(\mathbf{a}\cos\alpha - \mathbf{b} \right) + k\hat{\mathbf{\xi}}(\mathbf{a})\right].
\end{equation}
The difference between the price vectors is then
\begin{equation}\label{lim_cycle_diff}
\mathbf{b}-\mathbf{a} = (\cos\alpha-1)\mathbf{a} +\frac{k\cos\alpha}{1+(1-\hat{\gamma})\cos\alpha}\hat{\mathbf{\xi}}(\mathbf{a}).
\end{equation}
We are interested in the angle between these vectors, because it determines the range of variation of the price. We have scaled the price vectors to have a magnitude equal to one, so $\mathbf{a}\cdot\mathbf{b}=\cos\alpha$. Using this relationship and Walras' law, and taking the absolute square of both sides of Equation (\ref{lim_cycle_diff}) we find, after some rearrangement,
\begin{equation}\label{lim_cycle_angle}
(1-\cos^2\alpha)\left[ 1 + (1-\hat{\gamma})\cos\alpha \right]^2 = k^2\cos^2\alpha |\mathbf{\xi}(\mathbf{a})|^2.
\end{equation}
As the damping $\hat{\gamma}$ gets very large, the second factor on the left-hand side of this equation simplifies, and Equation (\ref{lim_cycle_angle}) becomes the approximate equation
\begin{equation}\label{lim_cycle_angle_largegamma}
(1-\cos^2\alpha)\hat{\gamma}^2\cos^2\alpha \simeq k^2\cos^2\alpha |\mathbf{\xi}(\mathbf{a})|^2,\quad \hat{\gamma} \gg 1.
\end{equation}
Rearranging this equation and using the Taylor series expansion for $\cos^2\alpha$ gives an approximate value for $\alpha$,
\begin{equation}\label{lim_cycle_angle_largegamma_alpha}
\alpha \simeq \pm\frac{k}{\hat{\gamma}} |\mathbf{\xi}(\mathbf{a})|,\quad \hat{\gamma} \gg 1.
\end{equation}
Unless the excess demand $|\mathbf{\xi}(\mathbf{a})|$ is very large, $\alpha$ approaches zero as $\hat{\gamma}$ increases. Therefore, at large damping the gap between two price vectors in a two-point cycle is limited to a small angle.

\section{Discussion}
The second-order t\^{a}tonnement process proposed in this paper has some advantages over other proposals for constructing convergent iterative price adjustment mechanisms. Convergence in the proposed process derives from heterogeneous responses to changing price signals rather than from added constraints on the excess demand function or arbitrary limits on the size of the price adjustment. The size of the damping term can vary with the market, being either strong, as for local food markets, or weak, as for the stock market. Under discrete-time dynamics, prices can cycle endlessly, even with damping. However, damping reduces the size of such cycles. Therefore the model allows for small shifts in prices from one time interval to the next, even in relatively stable markets.

The proposed mechanism can be thought of as an aggregate approximation to an actual market composed of individuals. Unlike representative agent models, the proposed mechanism assumes that people behave differently from one another, and the different behaviours, when aggregated over a large number of people, leads to the second-order and damped response to changes in excess demand and changes in price. We showed explicitly how a second-order dynamic can emerge when some sellers set their prices based on price movements in the previous period, while others set their price based on sales and inventory data. As observed in experimental settings {\cite{ostrom_rules_1994}}, when participants use a decision heuristic based on returns from the previous time step, the net result is a ``pulsing'', or cyclic, variation in the outcome; this is characteristic of a second-order dynamic. Alternatively, a second-order dynamic could emerge in a market with a few ``bulls'', or optimistic sellers, some ``bears'' or pessimistic ones, and a large number of ``sheep'', sellers who go along with the trend. A further possibility is that fatigue and disinterest gradually dissipate a repetitive cycle of price-setting.

\section{Conclusion}
This paper presents an alternative to classical t\^{a}tonnement. As in the classical, first-order mechanism, changes in price arise from nonzero excess demand. The proposed mechanism differs from the classical one in two ways: 1) the response is second-order in time, so prices gradually accelerate, rather than jump; 2) there is a damping term that opposes rapid changes in price, and slows them down. We showed how such a mechanism can arise when heterogeneous agents follow different simple rules to set prices.

For the proposed model we showed that in continuous time a deviation from a stable equilibrium will return to the equilibrium. In discrete time we showed that two-point cycles are possible, but the difference between prices, as measured by the angle between the two price vectors, shrinks as the damping increases. We conclude that a second-order dissipative price-setting mechanism is a promising alternative to the classical first-order process.

\bibliography{second_order_tatonnement_damping}

\end{document}